\documentclass[twocolumn,prl]{revtex4}
\usepackage{amsfonts}
\usepackage[T1]{fontenc}
\usepackage{amsmath,amsbsy,amssymb,graphicx}
\usepackage{times}

\def\beginABC{\begin{subequations}}
\def\endABC{\end{subequations}}
\let\mathbf=\boldsymbol
\let\dprod=\prod

\begin{document}

\title{{\Large Skyrmion Burst and Multiple Quantum Walk}\\
{\Large in Thin Ferromagnetic Films}}
\author{Motohiko Ezawa}
\affiliation{Department of Applied Physics, University of Tokyo, Hongo 7-3-1, 113-8656,
Japan }

\begin{abstract}
A giant Skyrmion collapses to a singular point by emitting spin waves in a
thin ferromagnetic film, when external mangetic field is increased beyond
the critical one. The remnant is a single-spin flipped (SSF) point. The SSF
point has a quantum diffusion dynamics governed by the Heisenberg model. We
determine its time evolution and show the diffusion process is a
continueous-time quantum walk. We also analyze an interference of two SSF
points after two Skyrmion bursts. Quantum walks for $S=1/2$ and $1$ are
exact solvable. The system presents a new type of quantum walk for $S>1/2$,
where a SSF point breaks into $2S$ quantum walkers. It is interesting that
we can create quantum walkers experimentally at any points in a magnetic
thin film, first by creating Skyrmions sequentially and then by letting them
collapse simultaneously.
\end{abstract}

\date{\today }
\maketitle
\affiliation{Department of Applied Physics, University of Tokyo, Hongo 7-3-1, 113-8656,
Japan }

Skyrmions are solitons in a nonlinear field theory, playing essential roles
in almost all branches of physics\cite{Skyrmion}. In particular, magnetic
thin films have recently attracted much attention owing to real-space
observations of Skyrmions\cite{Yu}. A Skyrmion crystal\cite{Mohlbauer} as
well as a single Skyrmion\cite{Yu} have been identified in chiral magnetic
thin films. In spite of their stability guaranteed topologically, an
intriguing feasibility arises that we are able to create them and destroy
them experimentally by breaking the continuity of the field. This is indeed
the case for giant Skyrmions in magnetic thin films\cite{EzawaGiant}. By
applying femtosecond optical pulse irradiation focused on a micrometer spot,
it is possible to destroy the magnetic order locally\cite{Ogasawara}. Then,
the dipole-dipole interaction (DDI) generates an effective magnetic field,
leading to a new magnetic order, that is a Skyrmion spin texture\cite%
{EzawaGiant}. On the other hand, as the magnetic field increases beyond the
critical one, the Skyrmion radius decreases and suddenly shrinks to zero by
emitting spin waves. This is the Skyrmion burst.

In this paper we investigate the fate of a Skyrmion after its burst. The
notion of the topological stability is lost when the radius becomes in the
order of the lattice constant. We consider the two-dimensional ferromagnet
with up spins. The spin at the Skyrmion center is precisely oriented
downward, which is topologically protected. The remnant of a Skyrmion burst
is expected to be a single-spin flipped (SSF) point, which is a single down
spin in the up-spin ferromagnet. It has a quantum diffusion dynamics
governed by the Heisenberg Hamiltonian. When the spin is $S=\frac{1}{2}$,
solving this problem exactly, we find that the down spin hops about
two-dimensional lattice points without canting. Namely, a SSF point can be
regarded as a two-dimensional continuous-time quantum walker. We have
verified manifestations of a quantum walk which differentiate from a
classical random walk: The probability density takes the maxima at moving
fronts and there appear oscillations between moving fronts. We then analyze
an interference of two SSF points after two Skyrmion bursts. The diffusion
dynamics is exactly solvable also for $S=1$. When the spin is higher ($S>%
\frac{1}{2}$), a SSF point breaks into $2S$ quantum walkers. This is a new
type of quantum walk, which we may call a multiple quantum walk. We are able
to create quantum walkers experimentally at any points in a magnetic thin
film, first by creating Skyrmions sequentially and then by letting them
collapse simultaneously.

A quantum walk is a quantum analogue of a classical random walk\cite%
{Aharonov,Fahri}. The quantum walk corresponds to the tunnelling of quantum
particles into several possible sites, generating large coherent
superposition states and allowing massive parallelism in exploring multiple
trajectories. The quantum walk is expected to have implications for various
fields, for instance, as a primitive for universal quantum computing and
systematic quantum algorithm engineering. Recently quantum walks have been
experimentally demonstrated using nuclear magnetic resonance\cite{Ryan},
trapped ions\cite{Zahr,Karski}, photons in fibre optics\cite{Schreiber} and
waveguides\cite{Peruzzo}. Our work presents an additional example of quantum
walks in magnetic thin films.

\textit{Giant Skyrmion:} We use the classical spin field of unit length, $%
\mathbf{n}=(n_{x},n_{y},n_{z})$, to describe the spin texture whose scale is
much larger than the lattice constant $a=0.3$[nm]. The Hamiltonian consists
of the anisotropic nonlinear sigma term $H_{J}$, the DDI term $H_{D}$ and
the Zeeman term $H_{Z}$, whose continuous versions read 
\begin{eqnarray}
H_{J} &=&\frac{1}{2}\Gamma \int d^{2}x[\left( \nabla \mathbf{n}\right)
\left( \nabla \mathbf{n}\right) -\xi ^{-2}\left( n_{z}\right) ^{2}],
\label{SigmaModel} \\
H_{D} &=&\frac{\Omega }{4\pi }\int d^{2}xd^{2}x^{\prime }\frac{\mathbf{n}(%
\mathbf{x})\cdot \mathbf{n}(\mathbf{x}^{\prime })}{|\mathbf{x}-\mathbf{x}%
^{\prime }|^{3}},  \label{DipolInter} \\
H_{Z} &=&-\Delta _{Z}\int \frac{d^{2}x}{a^{2}}\,n_{z}(\mathbf{x}),
\label{ZeemaInter}
\end{eqnarray}%
with the exchange energy $\Gamma $, the single-ion anisotropy constant $\xi $%
, the DDI strength $\Omega $, and the Zeeman energy $\Delta _{Z}$. The
ground state is the spin-polarized homogeneous state, $\mathbf{n}=(0,0,1)$,
under the field perpendicular to the plane.

The simplest cylindrical symmetric spin texture is%
\begin{align}
n^{x}(\mathbf{x})& =-\sqrt{1-\sigma ^{2}(r)}\cos \theta ,  \notag \\
n^{y}(\mathbf{x})& =-\sqrt{1-\sigma ^{2}(r)}\sin \theta ,\quad n^{z}(\mathbf{%
x})=\sigma (r),  \label{SkyrmField}
\end{align}%
with $\theta $ the azimuthal angle. The function $\sigma (r)$ should satisfy
the boundary conditions $\sigma (r)=-1$ at $r=0$ to avoid the
multivalueness, and $\sigma (r)\rightarrow 1$ as $r\rightarrow \infty $ to
approach the ground state. The spin texture (\ref{SkyrmField}) has the
Pontryagin number $Q_{\text{sky}}=1$, and hence it describes a Skyrmion. In
particular, there exists a disk-like spin texture(magnetic bubble domain)
with a domain wall at $r=R$. By changing variable as $\sigma (r)=\tanh \tau
\left( r\right) $, the equation of motion without the DDI term reads%
\begin{equation}
\frac{d^{2}\tau }{dr^{2}}+\frac{1}{r}\frac{d\tau }{dr}=\left[ \left( \frac{%
d\tau }{dr}\right) ^{2}-\left( \frac{1}{r^{2}}+\frac{1}{\xi ^{2}}\right) %
\right] \tanh \tau .
\end{equation}%
We solve it by setting $\left\vert \tanh \tau \right\vert =1$ in the
right-hand side, 
\begin{equation}
\sigma (r)=\tanh \left[ \log \left[ \frac{I_{1}\left( r/\xi \right) }{%
I_{1}\left( R/\xi \right) }\right] \right] ,  \label{SkyrmSigma}
\end{equation}%
with $I_{1}(x)$ the modified Bessel function. This is almost the exact
solution outside the domain wall. Actually it gives also an excellent
approximation to the domain wall. The excitation energy is given by\cite%
{EzawaGiant}%
\begin{equation}
E_{\text{sky}}(R)=\frac{4\pi \Gamma R}{\xi }-\Omega \lbrack R\ln \frac{R}{%
d_{F}}-R]+\pi \frac{R^{2}}{a^{2}}\Delta _{Z}+4\pi \Gamma ,
\label{SkyrmEnerg}
\end{equation}%
where the ground-state energy is subtracted, and $d_{F}$ is the thickness of
the film. The Skyrmion radius $R$ is determined by minimizing $E(R)$ with
respect to $R$. Since $R$ is as large as $1\mu $m for typical sample
parameters, we have called it a giant Skyrmion\cite{EzawaGiant}. The present
Skyrmion is stabilized dynamically by the competition among the DDI, the
Zeeman effect and the anisotropy of the film. The number of down spins is of
the order $(R/a)^{2}$, which is about $10^{7}$ for $R=1\mu $m.

According to a method of collective coordinate, the dynamics of the Skyrmion
radius is governed by the time-dependent Ginzburg-Landau equation,%
\begin{equation}
\frac{dR}{dt}=-L\frac{dE_{\text{sky}}(R)}{dR}=-L\left( \frac{4\pi \Gamma }{%
\xi }-\Omega \ln \frac{R}{d_{F}}+\frac{2\pi \Delta _{Z}}{a^{2}}R\right) ,
\label{TDGL}
\end{equation}%
where $L$ is the Onsager's constant. The Skyrmion is stable when $R$ is a
constant. It becomes unstable when the magnetic field increases beyond a
certain critical field. The DDI can stabilize the Skyrmion spin texture no
longer. In this case, solving (\ref{TDGL}), we find the Skyrmion radius to
change as%
\begin{equation}
R\left( t\right) =\left[ R_{0}+\frac{2a^{2}\Gamma }{\xi \Delta _{Z}}\right]
\exp \left[ -\frac{2\pi \Delta _{Z}L}{a^{2}}t\right] -\frac{2a^{2}\Gamma }{%
\xi \Delta _{Z}},
\end{equation}%
where $R_{0}$ is the initial radius. It shrinks to zero ($R=0$) after a
finite time interval,%
\begin{equation}
t=\frac{a^{2}}{2\pi \Delta _{Z}L}\log \left( \frac{\xi \Delta _{Z}}{%
2a^{2}\Gamma }R_{0}+1\right) .
\end{equation}%
It follows from (\ref{SkyrmSigma}) that, as $R\rightarrow 0$, $\sigma \left(
r\right) =1$ for $r>0$ and $\sigma \left( r\right) =-1$ for $r=0$. The spin
texture (\ref{SkyrmField}) becomes 
\begin{equation}
\mathbf{n}(0)=(0,0,1),\qquad \mathbf{n}(\mathbf{x})=(1,0,0)\quad \text{for }%
\mathbf{x}\neq 0.  \label{BlochPointCP}
\end{equation}%
The Skyrmion collapses into a singular point, which is a SSF point.

\textit{Skyrmion burst:} As the Skyrmion collapses, it emits spin waves:
About $10^{7}$ down spins are flipped upward, and the Skyrmion excitation
energy $E_{\text{sky}}(R_{0})$ is carried away, leaving only the energy $E_{%
\text{SSF}}$ of a SSF point.

To describe spin waves we parametrize the spin field as%
\begin{eqnarray}
n^{x}(\mathbf{x}) &=&\sigma (\mathbf{x}),\quad n^{y}(\mathbf{x})=\sqrt{%
1-\sigma ^{2}(\mathbf{x})}\sin \vartheta (\mathbf{x}),  \notag \\
n^{z}(\mathbf{x}) &=&\sqrt{1-\sigma ^{2}(\mathbf{x})}\cos \vartheta (\mathbf{%
x}),  \label{SpinField}
\end{eqnarray}%
so that the ground state is reached by setting $\sigma (\mathbf{x}%
)=\vartheta (\mathbf{x})=0$. Note that this parametrization is different
from that of the Skyrmion (\ref{SkyrmField}). The Lagrangian density is
given by%
\begin{equation}
\mathcal{L}=-\hbar S\sigma \dot{\vartheta}-\mathcal{H}\left( \sigma
,\vartheta \right) ,
\end{equation}%
where $S$ is the spin per atom. The Hamiltonian consists of the nonlinear
sigma model $H_{Z}$ and the Zeeman term $H_{Z}$. We may neglect the DDI $%
H_{D}$ to discuss spin waves. Substituting the above configuration into the
Hamiltonian density and taking the leading order terms in $\sigma $ and $%
\vartheta $, we obtain%
\begin{equation}
\mathcal{H}=\frac{\Gamma }{2}[\left( \nabla \sigma \right) ^{2}+\left(
\nabla \vartheta \right) ^{2}]+\left( \frac{1}{\xi ^{2}}+\frac{\Delta _{z}}{2%
}\right) (\sigma ^{2}+\vartheta ^{2}).
\end{equation}%
The Euler-Lagrange equations are easily written down. There exists the
outgoing propagating wave solution, 
\begin{subequations}
\begin{eqnarray}
\sigma \left( r,t\right) &=&J_{0}\left( kr\right) \sin \omega t-N_{0}\left(
kr\right) \cos \omega t, \\
\vartheta \left( r,t\right) &=&J_{0}\left( kr\right) \cos \omega
t+N_{0}\left( kr\right) \sin \omega t,
\end{eqnarray}%
up to an appropriate initial condition, where the dispersion relation reads 
\end{subequations}
\begin{equation*}
\hbar \omega =\frac{\Gamma }{S}\mathbf{k}^{2}+\frac{1}{S}\left( \frac{2}{\xi
^{2}}+\Delta _{z}\right) .
\end{equation*}%
The asymptotic behavior is $\sigma \left( r,t\right) =\sqrt{2/\pi kr}\sin
(kr-\omega t)$, etc., for $r\gg R$. It carries away spins and energies from
a Skyrmion.

\textit{Quantum walk:} The Skyrmion collapses into a singular point in the
ferromagnet. The remnant is a SSF point in a ferromagnet. The Skyrmion can
be no longer a classical object but a quantum object.

Our concern is about the dynamics of a SSF point after the skyrmion
collapse. We show that the dynamics is described as a two-dimensional
continuous-time quantum walk as governed by the anisotropic Heisenberg model.

The Heisenberg model is expressed as 
\begin{equation}
H_{J}=\frac{-J}{2}\sum_{\left\langle i,j\right\rangle
}(S_{i}^{+}S_{j}^{-}+S_{i}^{-}S_{j}^{+}+2S_{i}^{z}S_{j}^{z})-D\sum_{i}\left(
S_{i}^{z}\right) ^{2},  \label{HamilHeise}
\end{equation}%
from which the nonlinear sigma model (\ref{SigmaModel}) follows together
with $\Gamma =(1/2)S^{2}J$ and $\xi ^{-2}=4D/Ja^{2}$. The Zeeman term is $%
H_{Z}=-(\Delta _{Z}/S)\sum_{i}S_{i}^{z}$. The DDI is irrelevant. Here, the
index $i$ runs over the two-dimensional lattice points, $i\in \mathbb{Z}^{2}$%
. The ground state $|g\rangle $ is the up-spin polarized state defined by 
\begin{equation}
S_{i}^{z}|g\rangle =S|g\rangle ,\quad \quad S_{i}^{+}|g\rangle =0.
\end{equation}%
A SSF point is generated at $m=\left( m_{x},m_{y}\right) \in \mathbb{Z}^{2}$
by applying $(S_{m}^{-})^{2S}$ to the ferromagnetic ground state, 
\begin{equation}
\left\vert m\right\rangle =(S_{m}^{-})^{2S}\left\vert g\right\rangle ,
\end{equation}%
where only the spin at the lattice site $m$ is pointed downward.

First we study the case $S=\frac{1}{2}$ in details, since the problem is
exactly solvable. It is straightforward to show that%
\begin{eqnarray}
H\left\vert m\right\rangle &=&\frac{-J}{2}\sum_{p=\pm
1}(|m_{x}+p,m_{y}\rangle +|m_{x},m_{y}+p\rangle )  \notag \\
&&+\left( \frac{J}{2}+\Delta _{z}\right) \left\vert m\right\rangle .
\label{StepA}
\end{eqnarray}%
The energy of a SSF point is $E_{\text{SSF}}=\langle m|H|m\rangle $, which
is much smaller than the energy $E_{\text{sky}}(R_{0})$ of a Skyrmion.

The dynamics of a SSF point is governed by the Schr\"{o}dinger equation $%
i\hbar d\psi /dt=H\psi $. The time evolution reads $\left\vert \psi \left(
t\right) \right\rangle =e^{-i\frac{t}{\hbar }H}\left\vert 0\right\rangle $,
where $\left\vert 0\right\rangle $ denotes the initial state containing a
SSF point at site $m=(0,0)$. By applying the\ Hamiltonian to the state
containing one SSF point, the point remains at the same site or is shifted
to one of the neighboring sites. The resulting state is a coherent
superposition of them. It is notable that the spin is strictly parallel to
the $z$-axis: It never cants. The down spin diffuses by time evolution in
this way. We can regard a SSF point as a quantum walker.

The quantum diffusion process of a SSF point is governed by a
continuous-time quantum walk. Continuous-time quantum walk in one dimension
is studied in the refs.\cite{Konno,ben}. The process is a two-dimensional
extension of continuous-time quantum walk. In ref.\cite{Konno}, the
probability is obtained by the direct calculation of infinite multiplication
of matrix. In ref.\cite{ben}, Laplace transformation is used. We present a
new derivation using the generating function, which is an easy way to
generalize to a higher dimension.

\begin{figure}[t]
\centerline{\includegraphics[width=0.5\textwidth]{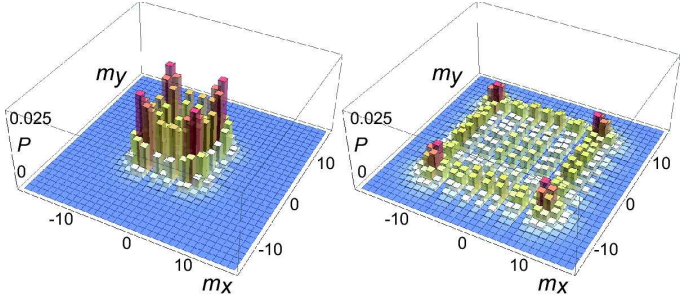}}
\caption{{}(Color online) (a) The probability dentity at $t=5/J$ and $t=10/J$%
. The horizontal axes are $m_{x}$ and $m_{y}$.}
\label{FigProb}
\end{figure}

Expanding the exponential, we write the time evolution as%
\begin{equation}
e^{-i\frac{t}{\hbar }H}|0\rangle \equiv \sum_{m}\mathfrak{C}_{t}\left(
m\right) |m\rangle .
\end{equation}%
Based on a combinatorial method known in the random walk theory, we can show
that the coefficient $\mathfrak{C}_{t}(m)$ is determined from the following
generating function,%
\begin{eqnarray}
&&\exp \left[ it\left\{ \frac{-J}{2}\left( \xi +\frac{1}{\xi }+\eta +\frac{1%
}{\eta }\right) +\left( \frac{J}{2}+\Delta _{z}\right) \right\} \right] 
\notag \\
&=&\sum_{m_{x},m_{y}}\mathfrak{C}_{t}(m)\xi ^{m_{x}}\eta ^{m_{y}},
\end{eqnarray}%
where $m=(m_{x},m_{y})$. It is determined in a closed form,%
\begin{equation}
\mathfrak{C}_{t}\left( m\right) =e^{it(\frac{J}{2}+\Delta
_{z})}i^{\left\vert m_{x}\right\vert +\left\vert m_{y}\right\vert
}J_{\left\vert m_{x}\right\vert }(Jt)J_{\left\vert m_{y}\right\vert }\left(
Jt\right) ,
\end{equation}%
with the use of the generating function of the Bessel function,%
\begin{equation}
\exp \left[ \frac{z}{2}\left( s-\frac{1}{s}\right) \right]
=\sum_{n}s^{n}J_{n}\left( z\right) .
\end{equation}%
The probability at site $m$ is given by $P_{t}(m)=|\mathfrak{C}_{t}(m)|^{2}$%
. The total probability is conserved, $\sum_{m}P_{t}\left( m\right) =1$, as
is easily checked based on the formula $\sum_{n}J_{n}^{2}\left( z\right) =1$.

We have thus solved the diffusion problem analytically. We show the
probability density as a function of site $m$ in Fig.\ref{FigProb}. The
probability density takes the maximum value not at the center but at the
fronts. Inside the maximum values, the probability density exhibits an
oscillatory behavior in the scale of lattice constant. They are
characteristic behavior of a quantum walk. This is highly contrasted to that
of a classical random walk, where the probability density is Gaussian with
the maximum value taking at the center and not oscillating.

\begin{figure}[t]
\centerline{\includegraphics[width=0.3\textwidth]{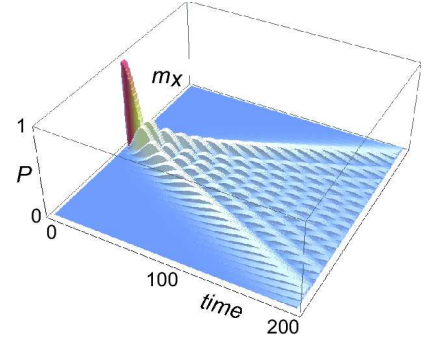}}
\caption{{}(Color online) (a) Time evolution of the probability dentity $%
P_{x}\left( t\right) $. The horizontal axes are the time $t$ and the site $%
m_{x}$.}
\label{FigEvol}
\end{figure}

The probability density is factorized as a direct product of the
probabilities along the $x$-axis and the $y$-axis, $%
P_{t}(m_{x},m_{y})=P_{t}(m_{x})P_{t}(m_{y})$, with 
\begin{equation}
P_{t}(m_{x})=\left\vert J_{|m_{x}|}\left( Jt\right) \right\vert ^{2}.
\label{ProbaDensi}
\end{equation}%
We illustrate the time evolution of $P_{t}(m_{x})$ in Fig.\ref{FigEvol}.
Long after the Skyrmion burst, it behaves asymptotically as%
\begin{equation}
P_{t}(m_{x})\simeq \frac{2}{\pi Jt}\cos ^{2}\left[ Jt-\frac{2\left\vert
x\right\vert +1}{4}\pi \right] .
\end{equation}%
The velocity of the front propagation is given by $v=2J/\pi $.

\begin{figure}[t]
\centerline{\includegraphics[width=0.3\textwidth]{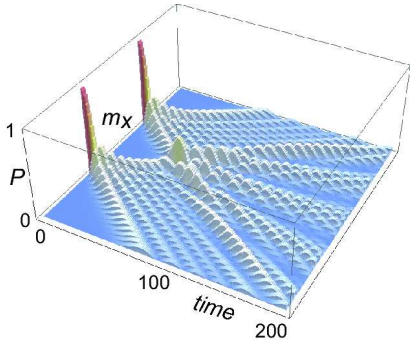}}
\caption{{}(Color online) (a) Time evolution of the probability dentity $%
P_{x}\left( t\right) $. The horizontal axes are the time $t$ and the site $%
m_{x}$.}
\label{FigInterF}
\end{figure}

We study interference effects of two Skyrmions bursts. We generate two
Skyrmions in a thin ferromagnetic film. By applying large magnetic field,
two Skyrmions collapse simultaneously and two SSF points are generated. They
are quantum mechanical objects, and the probability amplitude is given by a
linear superposition of each SSF points. The time evolution of these two SSF
points are shown in Fig.\ref{FigInterF}. An interference can be seen between
two SSF points. This is a manifestation of quantum mechanical properties of
a quantum walk.

\textit{Multiple quantum walk:} We discuss how the scheme is generalized to
the system where the spin is higher than $S=\frac{1}{2}$. In so doing, it is
adequate to recapitulate the diffusion process for $S=\frac{1}{2}$ from a
slightly different view point. For $S=\frac{1}{2}$, a SSF point at site $m$
is described by $|m\rangle =S_{m}^{-}|g\rangle $. When the Hamiltonian (\ref%
{HamilHeise}) acts on this state, the operator $S_{m}^{-}$ remains at the
same site or is shifted to one of the neighboring sites, as implied by (\ref%
{StepA}). Namely, the operator $S_{m}^{-}$ itself can be regarded as a
quantum walker.

When the spin is $S$, by generalizing the above picture, there are $2S$
quantum walkers at the initial SSF point, and then they diffuse. The state
is described by%
\begin{equation}
|m^{1},m^{2},\cdots ,m^{2S}\rangle =S_{m^{1}}^{-}S_{m^{2}}^{-}\cdots
S_{m^{2S}}^{-}|g\rangle ,  \label{GenerSite}
\end{equation}%
implying that a quantum walker is present at sites $m^{i}$. The number of $%
S_{m}^{-}$ is independent of time due to the conservation of the total $%
S^{z} $. Indeed, acting the Hamiltonian $H$ to the state (\ref{GenerSite}),
we find a coherent superposition of these states. The initial SSF point is
given by (\ref{GenerSite}) with $m^{i}=(0,0)$ for all $i$.

When $S=1$, there are two quantum walkers. It is easy to calculate $%
H|m^{1},m^{2}\rangle $ explicitly. The time evolution is also calculable.
Since the Clebsch-Gordan coefficient is constant, we obtain a concise
formula,%
\begin{equation}
\mathfrak{C}_{t}(m^{1},m^{2})=\exp [it\left( J-\Delta _{z}-D\delta
_{m^{1},m^{2}}\right) ]\mathfrak{C}_{t}(m^{1})\mathfrak{C}_{t}(m^{2}).
\end{equation}%
The probability density is factorizable, $P_{t}({m}^{1}{,m}%
^{2})=P_{t}(m^{1})P_{t}(m^{2})$, with (\ref{ProbaDensi}).

However, since the Clebsch-Gordan coefficient is not constant for $S\geq 
\frac{3}{2}$, 
\begin{equation}
S_{m}^{-}\left\vert q\right\rangle =\sqrt{S/2(S/2+1)-q(q+1)}\left\vert
q-1\right\rangle ,
\end{equation}%
the analysis is not simple. Each walker does not diffuse independently but
interacts each other. Nevertheless, to get a rough picture, we dare to
approximate $(S_{m}^{-})^{q}|q\rangle =C|q-1\rangle $ with a certain
constant $C$. Then, we find%
\begin{equation}
P_{t}(m^{1},m^{2},\cdots ,m^{2S})\cong \dprod\limits_{q=1}^{2S}P_{t}(m^{q}).
\end{equation}%
We may regard the diffusion process as a diffusion of $2S$ independent
walkers in this approximation.

We have studied the diffusion process at zero temperature. One might wonder
if it would survives spin-wave excitations. There exist no problem because
spin-wave excitations have a large gap induced by the anisotropy and the
external magnetic field.

We have obtained an analytical solution of the time evolution dynamics of
the SSF point generated by a Skyrmion burst. Its quantum diffusion process
is a spreading oscillatory propagating wave and shows an interference. This
is highly contrasted compared to a classical diffusion process, where the
dynamics is described by a Gaussian. Our system provide a new approach to
investigate properties of quantum walk.

I am very much grateful to N. Nagaosa for fruitful discussions on the
subject. This work was supported in part by Grants-in-Aid for Scientific
Research from the Ministry of Education, Science, Sports and Culture No.
22740196 and 21244053.

\end{document}